\documentclass[sigconf]{acmart}

\usepackage{mathtools}
\usepackage{multicol}
\usepackage{graphicx}
\usepackage{epsfig}
\usepackage{multirow}
\usepackage{dsfont}
\usepackage{bm}
\usepackage{ upgreek }
\usepackage{amssymb}
\usepackage{amsmath}

\usepackage{caption}
\usepackage{subcaption}

\def\BibTeX{{\rm B\kern-.05em{\sc i\kern-.025em b}\kern-.08emT\kern-.1667em\lower.7ex\hbox{E}\kern-.125emX}}
    
\copyrightyear{2019} 
\acmYear{2019} 
\setcopyright{acmcopyright}
\acmConference[UMAP '19]{27th Conference on User Modeling, Adaptation and Personalization}{June 9--12, 2019}{Larnaca, Cyprus}
\acmBooktitle{27th Conference on User Modeling, Adaptation and Personalization (UMAP '19), June 9--12, 2019, Larnaca, Cyprus}
\acmPrice{15.00}
\acmDOI{10.1145/3320435.3320464}
\acmISBN{978-1-4503-6021-0/19/06}

\begin{document}

%
\title[Power of the Few: Analyzing the Impact of Influential Users]{Power of the Few: Analyzing the Impact of Influential Users in Collaborative Recommender Systems}

\fancyhead{}

%
\author{Farzad Eskandanian}
\email{feskanda@depaul.edu}
\affiliation{%
  \institution{DePaul University}
  \city{Chicago}
  \state{Illinois}
  \country{USA}
}

\author{Nasim Sonboli}
\email{nasim.sonboli@colorado.edu}
\affiliation{%
  \institution{Colorado University}
  \city{Boulder}
  \state{Colorado}
  \country{USA}
}

\author{Bamshad Mobasher}
\email{mobasher@cs.depaul.edu}
\affiliation{%
 \institution{DePaul University}
 \city{Chicago}
 \state{Illinois}
 \country{USA}}

\begin{abstract}
Like other social systems, in collaborative filtering a small number of ``influential'' users may have a large impact on the recommendations of other users, thus affecting the overall behavior of the system. Identifying influential users and studying their impact on other users is an important problem because it provides insight into how small groups can inadvertently or intentionally affect the behavior of the system as a whole. Modeling these influences can also shed light on patterns and relationships that would otherwise be difficult to discern, hopefully leading to more transparency in how the system generates personalized content. In this work we first formalize the notion of ``influence'' in collaborative filtering using an {\it Influence Discrimination Model}. We then empirically identify and characterize influential users and analyze their impact on the system under different underlying recommendation algorithms and across three different recommendation domains: job, movie and book recommendations. Insights from these experiments can help in designing systems that are not only optimized for accuracy, but are also tuned to mitigate the impact of influential users when it might lead to potential imbalance or unfairness in the system's outcomes.
\end{abstract}


%

\maketitle

\section{Introduction}

Personalized recommender systems have become essential tools for users to navigate multitude of choices in large information or product spaces. One of the most commonly used approaches to personalized recommendation is Collaborative Filtering. The main tenet of this approach is to recommend items of interest to a user based on the preferences of other similar users in the system. Because of the social nature of these systems, a small group of ``influential'' users can have a significant impact on the behavior of the system towards other users. This type of influence may, in some cases, result in undesirable effects such as bias toward certain items, lack of diversity or imbalance in recommendations, and even potential security concerns such as making it easier to deliberately manipulate the system outcomes. 

This type of behavior is not uncommon in social networks and many studies in social network analysis have focused on how certain influential users may significantly affect the way in which information is propagated across such networks. Some studies have tried to extend this work in social networks to the realm of collaborative recommendation. There has also been much work in analyzing different ways collaborative systems can be manipulated to promote or demote items, for example by manipulating the behavior of certain ``power'' users. Overall, however, many aspects of this problem in the context of recommender systems have remained unexplored. For example, there has not been adequate exploration of what constitutes an influential user and how specifically such users, individually or in groups, impact other users in the system. There has also been little empirical analysis of relationships among influential users and between them and other users of the system. Finally, much of the previous work in this area does not extend beyond collaborative recommendation based on the standard k-Nearest-Neighbor (kNN) approach to more recent approaches such as matrix factorization. 

Our goal in this paper is to revisit the notion of influential users in recommender system and provide a more extensive analysis of their characteristics and their impact on other users under different conditions, including when different algorithms (such as matrix factorization) are used as well as across different recommendation domains. 

We are particularly interested in addressing three main research questions:

\begin{itemize}
    \item \textbf{RQ1}: To what degree do individual users influence the recommendation lists of other users?
    
    \item \textbf{RQ2}: What are the similarities among the most influential users in the system and their relationships to other users?
    
    \item \textbf{RQ3}: What are the most important characteristics that make a user highly influential?
\end{itemize}

To answer these questions, we propose an {\it Influence Discrimination Model} which measures user $u$'s influence based on how recommendation lists of other users change if $u$ is removed from the system. Knowing the influence of each user on other users' recommendations, we show the impact a few users can have. We also identify a set of factors that can be used as a set of features, in conjunction with the Influence Discrimination Model, to characterize the most influential users and to develop a prediction model for more efficiently and accurately estimating influence of a given user.

In our experiments we focus on two recommendation algorithms: Non-negative Matrix Factorization (NMF) and User-based Nearest Neighbor (User-KNN), which are two of the most commonly used Collaborative Filtering approaches. We conduct several experiments designed to address the aforementioned research questions and we analyze the results of these experiments for each of these algorithms across three different data sets. In particular, we show how the choice of recommendation algorithm and its parameters can have a significant impact on the size and the scope of impact exerted by top influential users. We also make some observations about how the characterization of influential users and their relationships to other users may be different under different algorithms.

\section{Related Work}
\label{sec:related_work}

\textbf{Influence Maximization in Social Networks.}

The notion of centrality of a user in Social Network Analysis (SNA) has been studied extensively \cite{bhagat2012maximizing, kempe2003maximizing, cha2010measuring}. There have been some effort to adapt this work in SNA to other domains including collaborative recommender systems \cite{lathia2008knnSNA}. 
The work in social network analysis requires an explicit social network. The notion of influence in the context of viral marketing was first introduced by Domingos, et al. \cite{domingos2001mining} as the expected lift in profit with the goal of maximizing it. They used a network of users in which the relations are defined based on the similarity of users in terms of rated items in their profiles. In the context of collaborative recommendation, the underlying network is also implicit and is again induced based on similarity relationships among users. Our work is most closely related to the RecMax approach proposed by Goyal and Lakshmanan \cite{goyal2012recmax}. RecMax tries to adapt the notion of influence maximization from social network analysis to recommender systems. The main idea is to maximize the propagation of a new item through recommendations by injecting the item into the profiles of a group of users which they call seed users. Therefore, their problem is to find a group of users who can maximize the spread of a new item through recommendations. In contrast to this objective, our work is aimed at studying the influence of individual users or small sets of users on recommendations measured in terms of the amount of change in the recommendation lists caused by the existence of a user in the dataset.
\\

\noindent\textbf{Power User Attacks in Recommender Systems.}

Much work has focused on profile injection attacks in collaborative filtering \cite{mobasher2007toward}, the idea that bogus user profiles can be added in order to manipulate predicted ratings, thus promoting or demoting recommended items. In particular, some prior work has considered Power User Attacks (PUA) \cite{wilson2013power, seminario2014assessing}. Power users are those who might have out-sized influence on predicted ratings and thus might be able to manipulate the behavior of the system by providing false ratings for one or more items. The impact of power users is typically measured by the shift in ratings that power users can cause on selected new items. 

Our work is different in many respects from the prior work on PUAs. First, the prior work on PUAs generally assumes that there is a potential harm or attack that power users can cause in recommender systems. We do not study such vulnerabilities or attacks, instead we study the question of what percentage of users exert their influence through recommendations and to what extent they do that. We do not see their influence as attacks and we do not study how they can promote a new item. Secondly, three heuristics are used for identifying power users: in-degree centrality, aggregated similarity score, and number of ratings. In contrast, we study the characteristics which enable a user to have a greater influence on the recommendations compared to other users including the heuristics they have used. In other words, our analysis can lead to designing more accurate heuristics for estimating the influence of a user but the goal of our research is completely different from the notion of PUA and its purpose. We also use our Influence Discrimination Model as a more accurate ground truth for training a predictive model.
\\

\noindent\textbf{Measures of Influence in Rating-based Systems.}

There has been some prior work on measuring influence in ratings-based systems and on characterizing influential users. The work by Rashid et al. \cite{rashid2005influence} is the closest to our work and partly inspired our research. This work is the first attempt to study the influence of users based on various characteristics such as the number of ratings in a user profile, popularity of rated items by that user, etc. However, there are many aspects to this problem that are missing in this work. For example, in their experiments they only used one dataset (MovieLens 1M) and the recommendations are limited to neighborhood-based methods. Also, their notion of user influence is based on the shift in ratings caused by the existence of a user profile in the dataset. We extend this work to other recommendation methods such as matrix factorization and other data sets. Also, we introduce a more accurate notion of influence using our Influence Discrimination model which is based on the ranking measure of difference in recommendation lists. We explain this difference in more details in Section \ref{IDM model}. Finally, We provide a more in depth analysis of the impact of influential users, visualize their characteristics, and analyze their relationships to other users, an aspect that has not been explored in earlier works. 

\section{Background and Definitions}

Let $\mathcal{U} = \{u_1,...,u_n\}$, and $\mathcal{I} = \{i_1,...,i_m\}$ be the set of users and items, respectively. The rating assigned by user $u$ to item $i$ is denoted by $r_{u,i}$. Also, the set of items rated by user $u$ is denoted by $I_u$. In ranking-based recommender systems, the goal is to rank all the items  for each user based on their predicted ratings and select the top-$l$ items for recommendation to $u$. We denote the recommendation set to user $u$ by $\mathcal{R}_u$.

One of the most common approaches used in collaborative filtering is based on the $k$-Nearest Neighbors rating prediction. The top-$k$ most similar users to $u$ are the $k$-Nearest Neighbors of $u$, denoted by $\mathcal{N}_u$. The estimated rating that user $u$ would assign to item $i$ can be computed as follows:
\begin{equation}
    r'_{u,i} = \frac{\sum_{v \in \mathcal{N}_u}{\sigma_{u,v} \hspace{0.3em} r_{v,i}}}{\sum_{v \in \mathcal{N}_u}{|\sigma_{u,v}|}}
\end{equation}
where $\sigma_{u,v}$ is a function measuring the similarity between the similarity between users $u$ and $v$. If an item is not rated by neighbors of $u$ we assign the average rating of that item among all the users as its predicted rating.

User $u$'s recommendation list $\mathcal{R}_u$ is consist of the top-$l$ items (which are not rated by $u$) according to the estimated ratings of all the items $i\in \mathcal{I}$:
\begin{equation}
    \mathcal{R}_u = \operatorname*{args\,max}_{i \in \mathcal{I}}^{l}{\hspace{0.3em} r'_{u,i}}
\end{equation}

An alternative approach to rating prediction is Matrix Factorization which is a shared latent space between users and items derived from the sparse rating matrix $M_{n\times m}$ containing users as rows and items as columns. Matrix Factorization is known for its scalability and accuracy of rating predictions. In this work, to find approximate factorization of $M \approx p.q^T$, we use Non-negative matrix factorization \cite{cichocki2009_NMF}. The objective in this method is to minimize the Euclidean distance between the approximated matrix $M' = p.q^T$ and the actual matrix $M$. More precisely, NMF minimizes $|| M - p.q^T ||_F^2$ with respect to $p$ and $q$, subject to $p,q \geqslant 0$. Using $M'$, user $u$'s recommendation list is estimated as:
\begin{equation}
    \mathcal{R}_u = \operatorname*{args\,max}_{i \in \mathcal{I}}^{l}{\hspace{0.3em} M'_{u,i}}.
\end{equation}

This objective function is a simplified version of the original NMF. Interested reader can refer to \cite{cichocki2009_NMF, fevotte2011_NMF_algorithms} for more details.

\section{Influence Discrimination Model} \label{IDM model}

Among these two recommendation approaches the recommendations of User-kNN are easier to interpret, because the predicted ratings are calculated based on similarities between users. Using the similarity function $\sigma$ between the target user $u$ and another user $v$, we can determine the exact contribution by user $v$ to each predicted rating $r'_{u,i}$. However, in matrix factorization because of the latent space transformation it is difficult to directly estimate the contribution of each user to predict ratings of another target user. While here we focus on User-kNN and Matrix Factorization, it should be obvious that similar challenges might exist in measuring the influence of users on predicted ratings in other approaches to collaborative filtering.

In order to develop a general definition of a user's influence on predicted ratings of other users, we propose the {\it Influence Discrimination Model}. In this model we define a user $u$'s influence in terms of $u$'s impact on the recommendation lists of other users in the system. To measure influence by a user $u$ we compare the generated recommendation lists for each user $v$ before an after removing $u$ from the system. More specifically, user $u$'s influence is defined by:

\begin{equation}
    influence(u) = \sum_{v \in \{\mathcal{U} \setminus u\}}{Jaccard Dist(\mathcal{R}_v^{(u)},  \overline{\mathcal{R}}_v^{(u)})}
\end{equation}
Where $\mathcal{R}_v^{(u)}$ is the top-$l$ recommendation list generated for user $v$ with user $u$'s ratings are included in generating recommendations, and $\overline{\mathcal{R}}_v^{(u)}$ is the top-$l$ recommendation list generated excluding user $u$. 
This normalized difference between recommendation lists by including and excluding $u$ captures the essential influence of $u$ on all the system's recommendations as a whole. 

Note that while this definition of influence can be used in general regardless of which underlying recommendation algorithm is used, the influence values computed for each user are, in fact, dependent on the choice of algorithm and its parameters. 

A variation of this model was first introduced by \cite{rashid2005influence} was measured by computing the average rating prediction shift on all items. However, there is a drawback in using prediction shift to measure influence. Usually, the number of items in $\mathcal{R}_u$ is small (e.g., less than 20). This means that the recommendation algorithm should be able to accurately model the preferences of a user such that it could retrieve a handful of relevant items among thousands or sometimes millions of items. This consideration is especially important when we are analyzing the output of recommendations, and what is sensible to users. Therefore, the shift in predicted ratings of items which do not appear in recommendation lists are insignificant. In our approach, however, a shift in predicted ratings only affects the influence computation if it results in certain items being added or dropped from recommendation lists of one or more users.

\section{Experiments and Analysis}

In this section we discuss our experimental methodology and our findings across three different data sets.

\subsection{Datasets} \label{Datasets_subsection}

In our analysis we used two publicly available datasets. The first dataset is MovieLens 1M\footnote{https://grouplens.org/datasets/movielens} (ML), a specific dataset for movie recommendation which is widely used in Recommender system's domain. ML contains 6,040 users and 3,702 movies and 1M ratings. Sparsity of ratings in this dataset is about 96\%.

The second dataset is Xing job recommendation from RecSys Challenge 2017\footnote{http://www.recsyschallenge.com/2017}. We selected the region with smallest number of jobs and randomly sampled 1,527 users. This sample contains 35,202 jobs with 96,035 ratings and rating sparsity of 99.8\%. The ratings in Xing dataset are specified by the importance of the user interaction with the recommended job. For example, if a user clicked on the reply button or application form of the recommended job that indicates the highest interest of user in that job and we set the highest rating for this interaction type.

The third dataset is a sample of the Book Crossing (BX) dataset. We randomly sampled 4,100 users from the original dataset mainly due to its large size. The sampled dataset contains 15,000 books and the sparsity of ratings is about 99.5\%.

\begin{table}[h]
  \centering
  \begin{tabular}{c c c c c}
    Method& Dataset& factors/k& $P@10$& $R@10$\\ [0.5ex]
    \hline
    NMF& ML& f=40& 0.26& 0.18\\
    NMF& Xing& f=70& 0.34& 0.40\\
    NMF& BX& f=150& 0.02& 0.04\\
    \hline
    kNN& ML& k=60& 0.27& 0.20\\
    kNN& Xing& k=10& 0.35& 0.42\\
    kNN& BX& k=60& 0.03& 0.05\\
    \hline 
  \end{tabular}
  \caption{Accuracy results}
  \label{tab:accuracy}
\end{table}

\subsection{Experimental Setup}

We ran specific experiments designed to address each of the aforementioned research questions, RQ1-RQ3 discussed in the Introduction. We used the Influence Discrimination model that we introduced in subsection \ref{IDM model} to identify influential users and analyze their characteristics. 
In the remainder of this section we discuss the details of each of these experiments and present our results. 

For our experiments we focused on the two recommendation methods (User-kNN and NMF) that we presented earlier. In each case, recommendations were generated using hyper-parameters that empirically resulted in highest accuracy measured in terms of precision and recall for top-$k$ recommendations which are denoted by $P@k$ and $R@k$, respectively \cite{herlocker2004evaluating}. In our experiments we selected the top-10 results as the recommendation lists for target users. Table \ref{tab:accuracy} shows the hyper-parameters used to obtain the best accuracy in each of the two recommendation methods, NMF and user-kNN, on various datasets. The main parameter that we have focused on for tuning is the number of factors in NMF and number of neighbors $k$ in user-kNN. 

\subsection{Scope and Impact of Influential Users}

\begin{figure*}[ht]
\centering
\begin{subfigure}{.33\textwidth}
  \centering
  \includegraphics[width=1\linewidth]{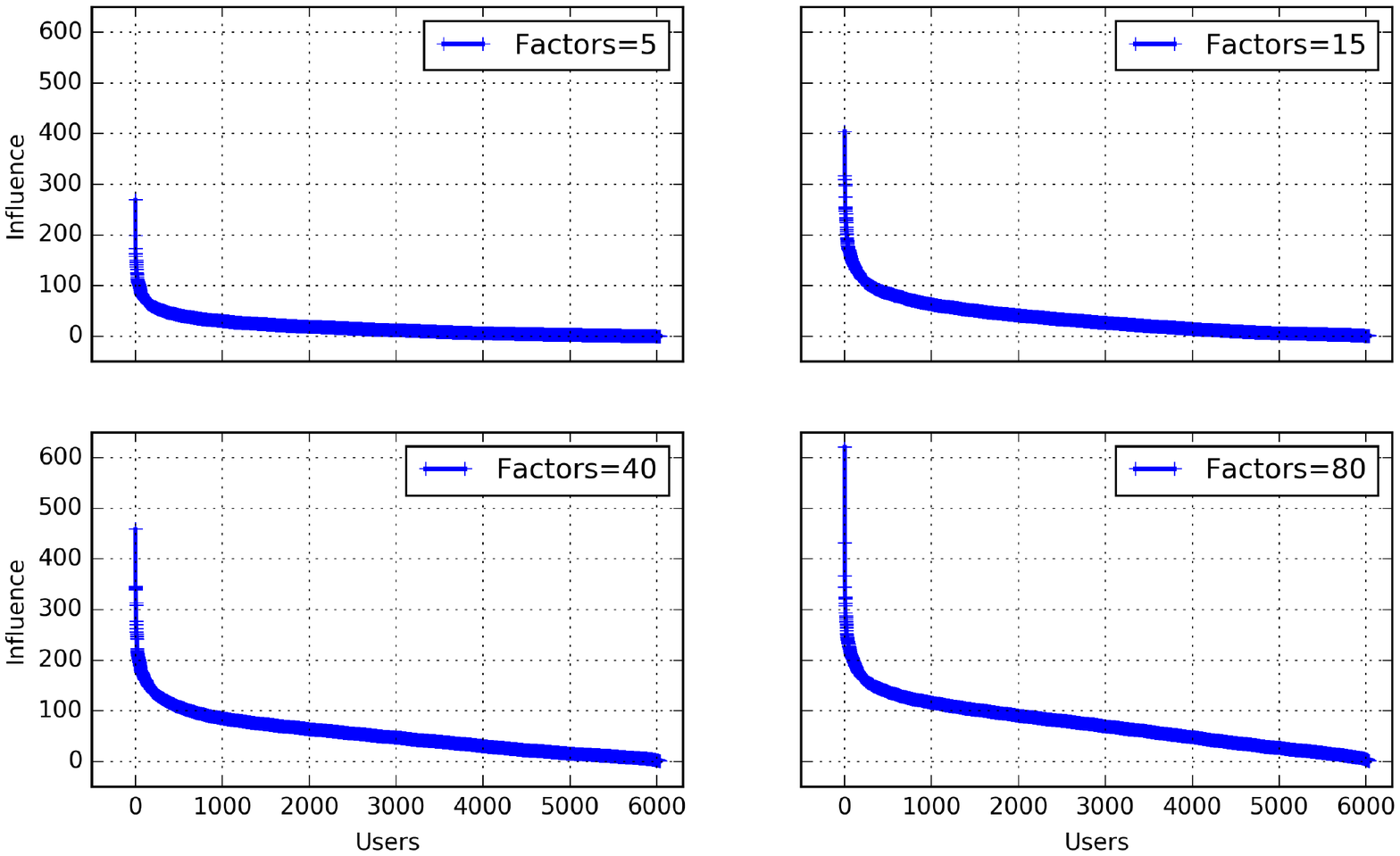}
  \caption{ML}
  \label{fig:Influencers_ML_NMF}
\end{subfigure}%
\begin{subfigure}{.33\textwidth}
  \centering
  \includegraphics[width=1\linewidth]{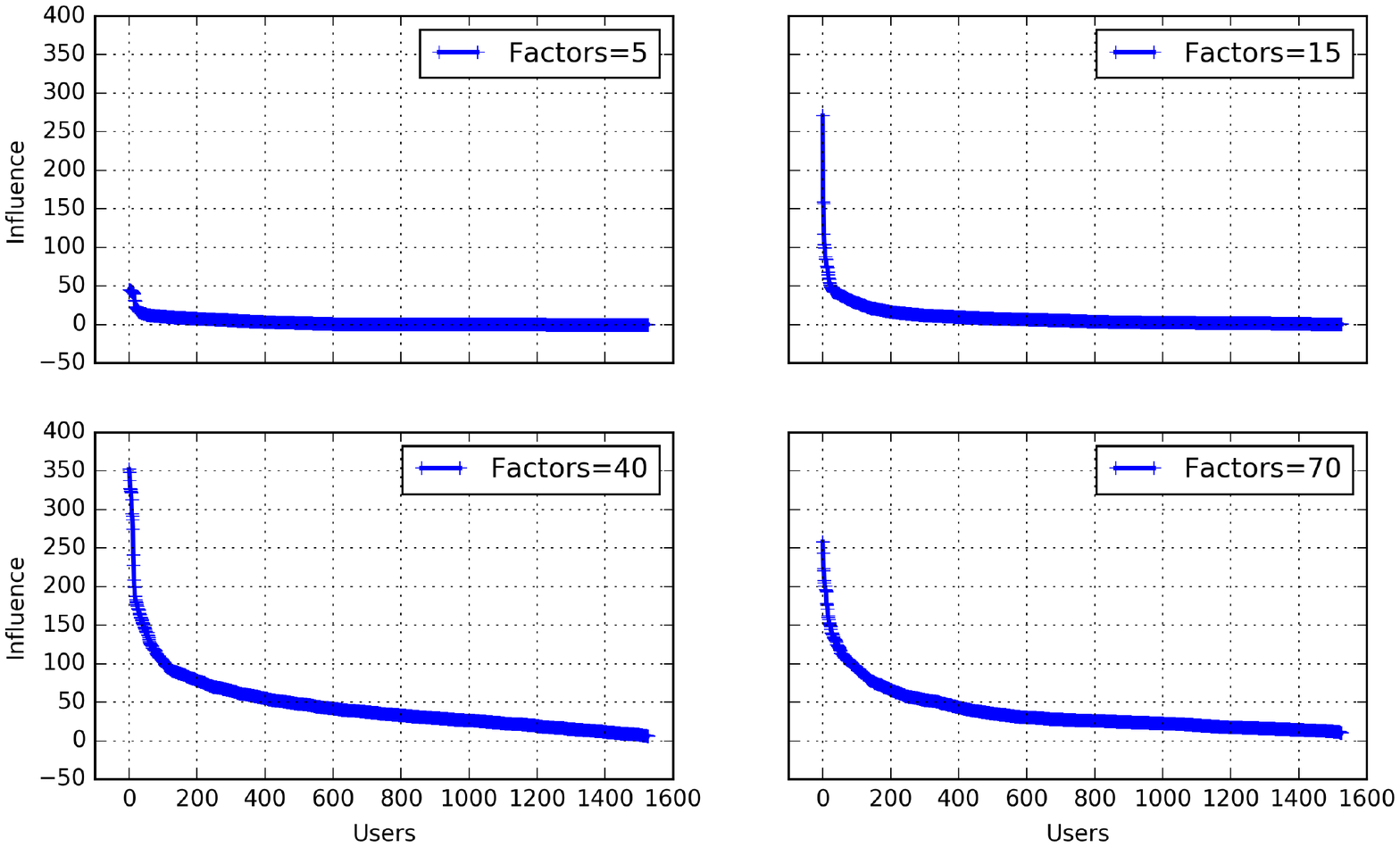}
  \caption{Xing}
  \label{fig:Influencers_XG_NMF}
\end{subfigure}%
\begin{subfigure}{.33\textwidth}
  \centering
  \includegraphics[width=1\linewidth]{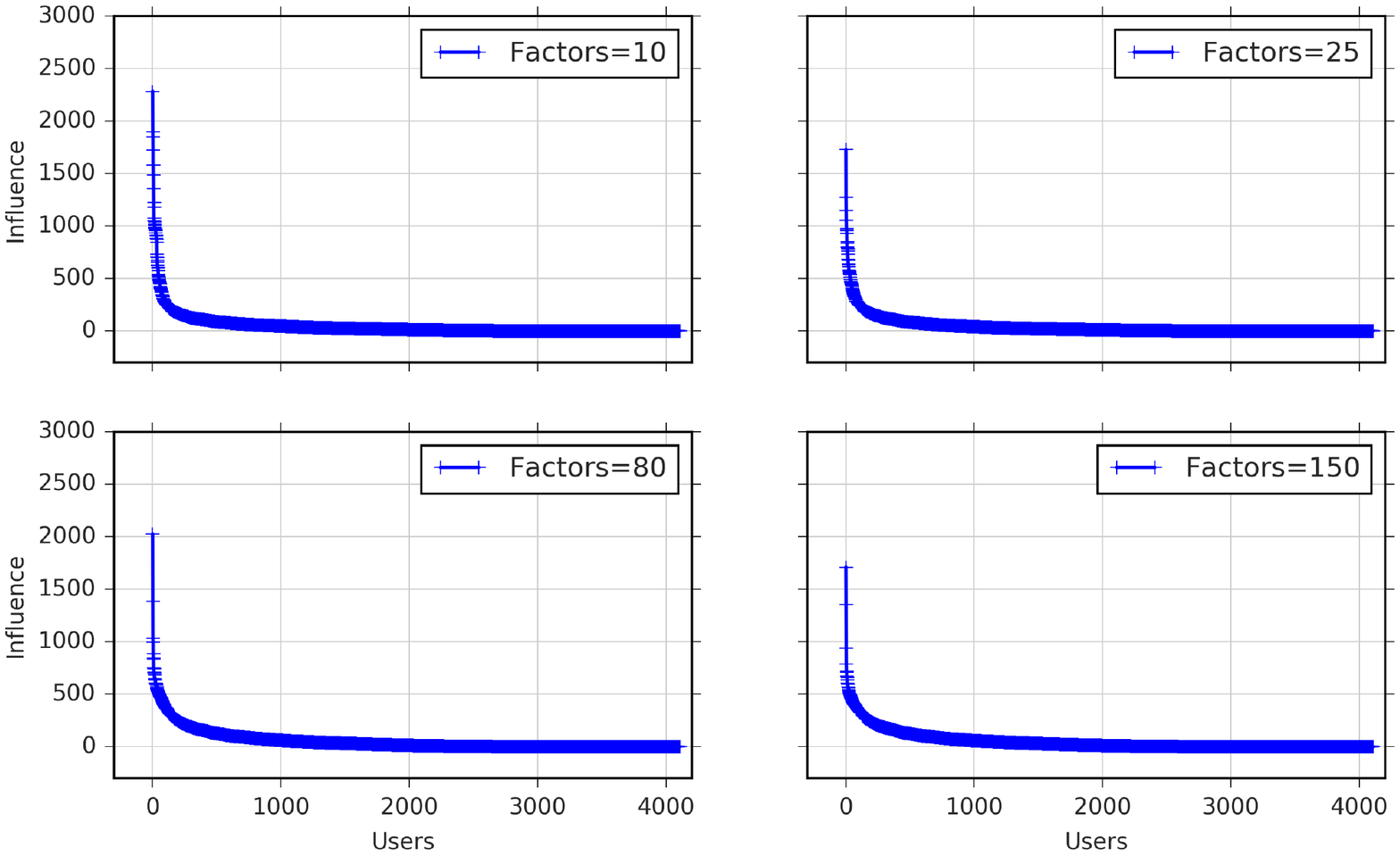}
  \caption{BX}
  \label{fig:Influencers_BX_NMF}
\end{subfigure}%
\caption{Influence of users on NMF's recommendations.}
\label{fig:influencers_NMF}
\end{figure*}

\begin{figure*}[ht]
\centering
\begin{subfigure}{.33\textwidth}
  \centering
  \includegraphics[width=1\linewidth]{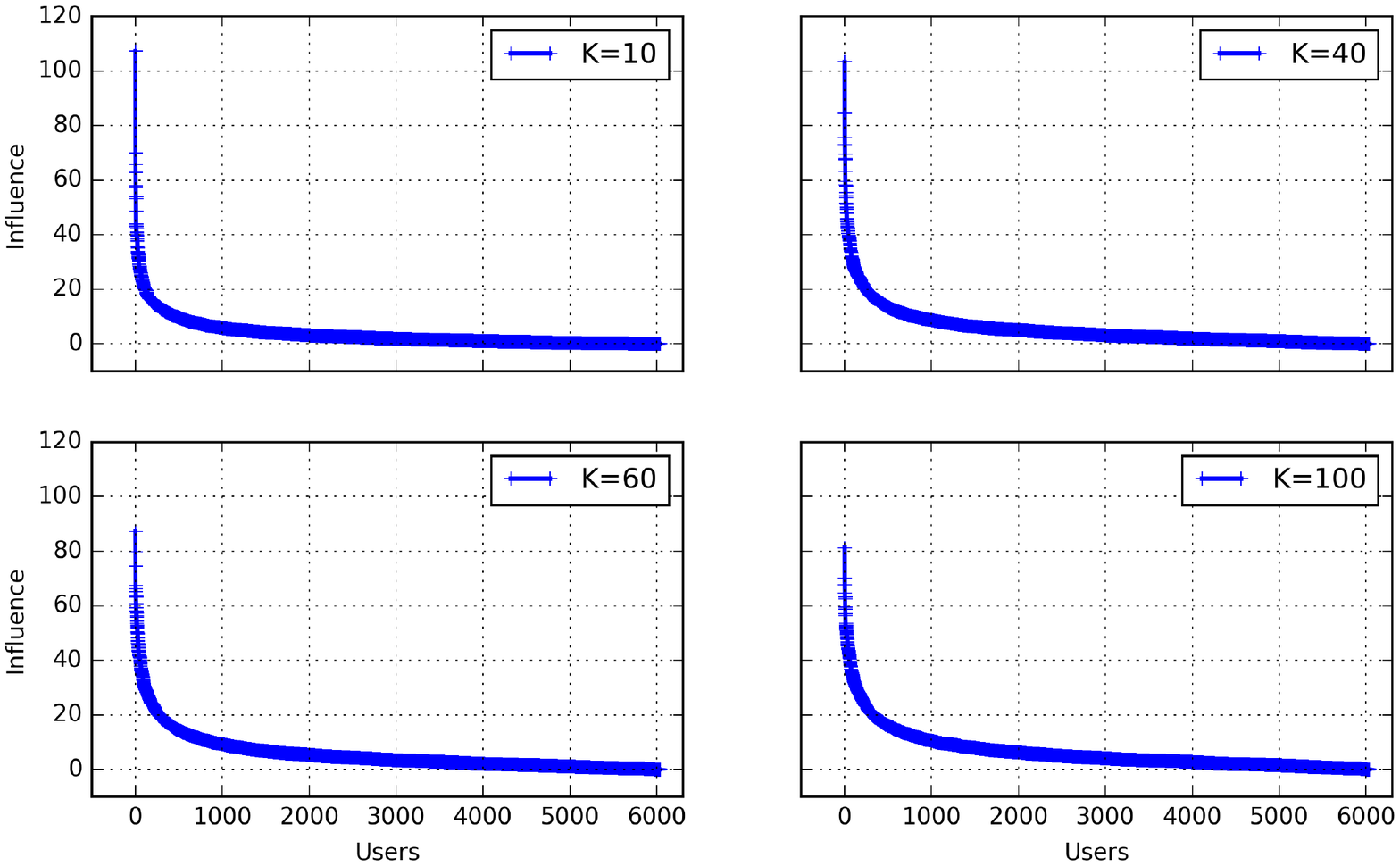}
  \caption{ML}
  \label{fig:Influencers_ML_kNN}
\end{subfigure}%
\begin{subfigure}{.33\textwidth}
  \centering
  \includegraphics[width=1\linewidth]{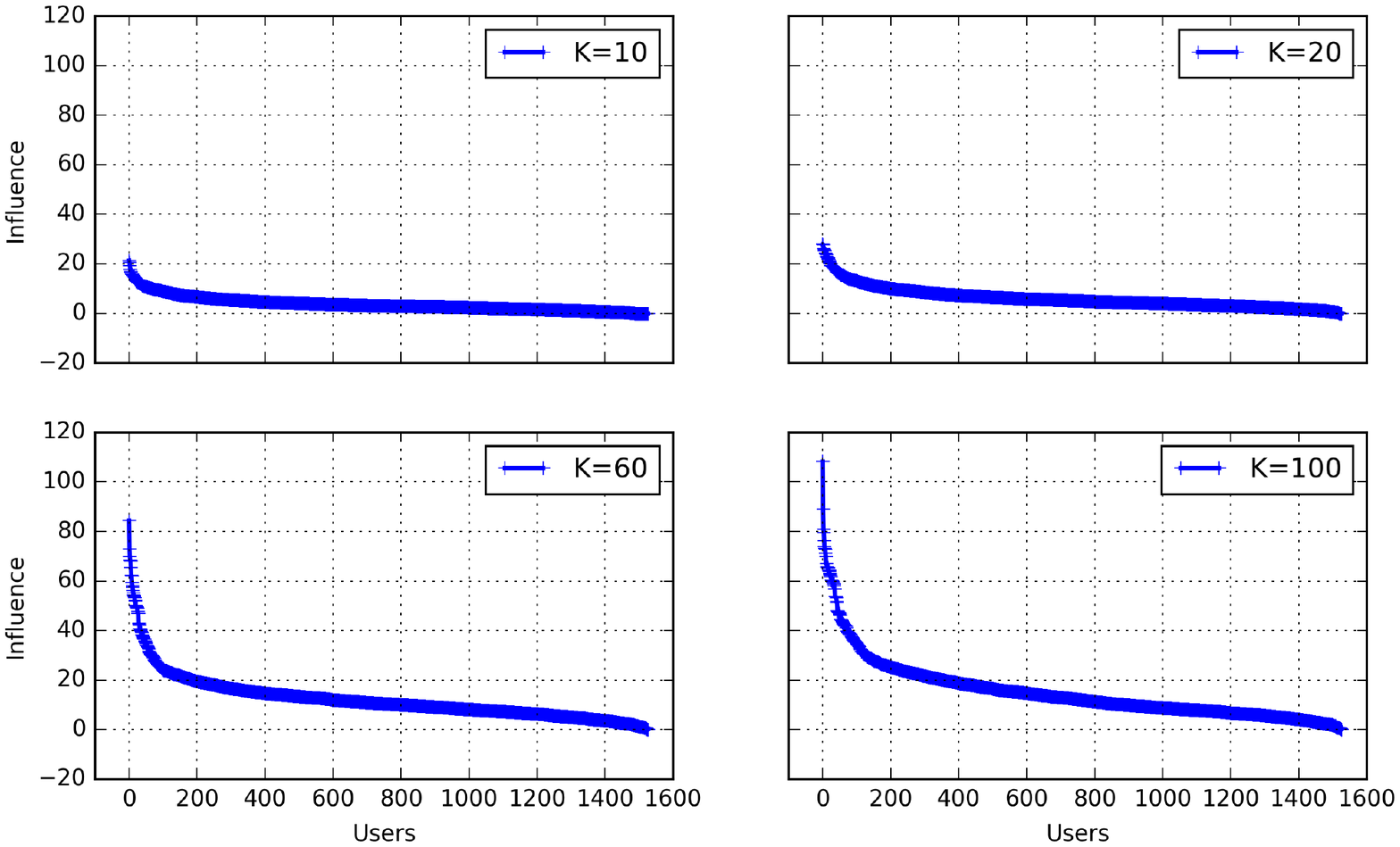}
  \caption{Xing}
  \label{fig:Influencers_XG_kNN}
\end{subfigure}
\begin{subfigure}{.33\textwidth}
  \centering
  \includegraphics[width=1\linewidth]{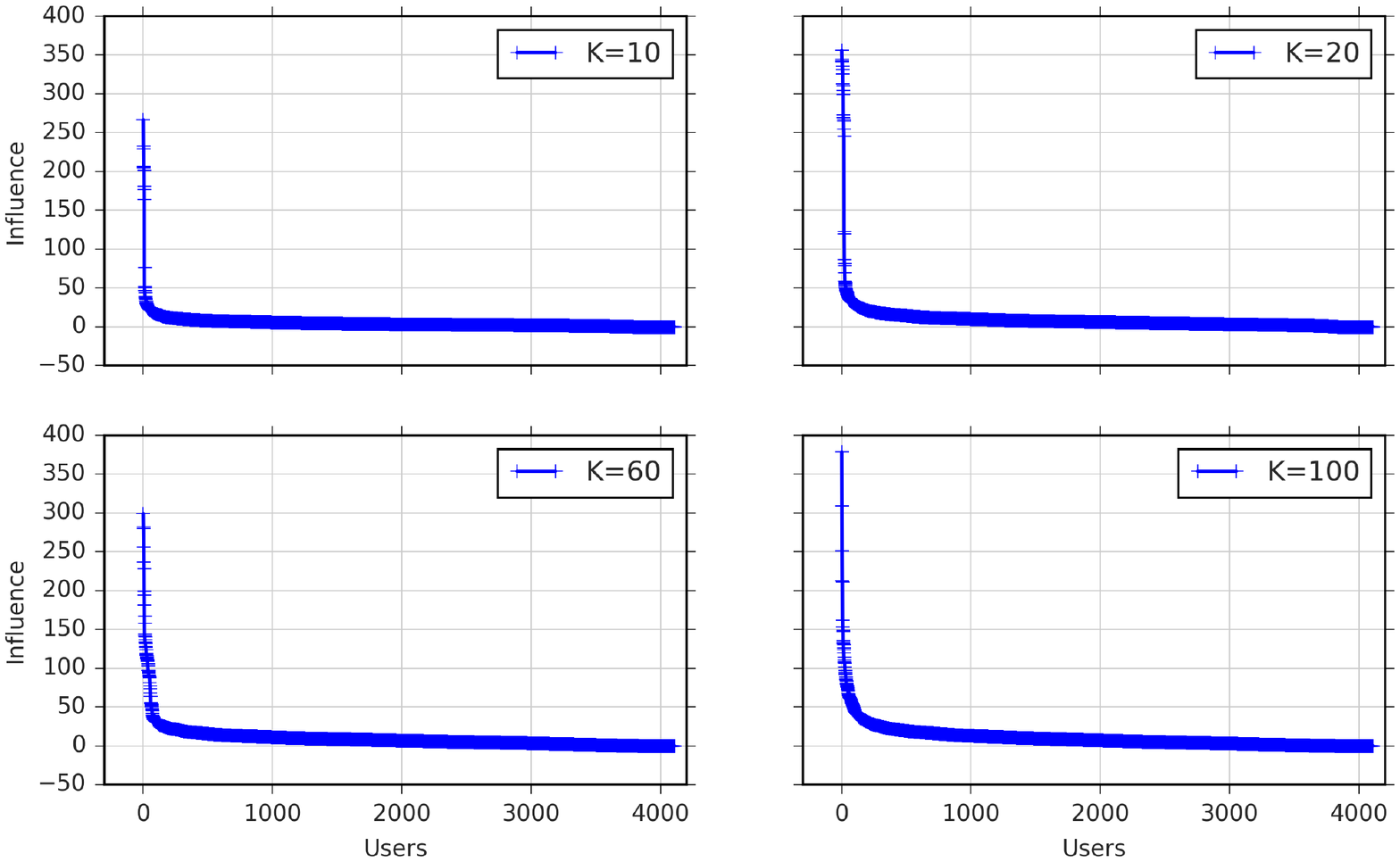}
  \caption{BX}
  \label{fig:Influencers_BX_kNN}
\end{subfigure}
\caption{Influence of users on the User-kNN's recommendations.}
\label{fig:influencers_kNN}
\end{figure*}

The first set of experiments where designed to address RQ1, i.e., to determine the distribution of influential users in the data and the degree to which they exert influence on other users. To this end we tried to visualize the distribution of influence values across users computed using the Influence Discrimination Model presented earlier. 
In each plot we ranked the users in the given data set based on their influence values. We varied hyper-parameters for each algorithm (the number of latent factors for NMF and the number of neighbors for User-kNN) to observe their impact on influence values.  Figures \ref{fig:influencers_NMF} and \ref{fig:influencers_kNN} show the distribution plots of influence values for the User-kNN and matrix factorization for a choice of hyper-parameter values across our three data sets. 

In general, these results give credence to the hypothesis that influential users follow a long-tail distribution: a small number of users are highly influential while the vast majority do not exert a lot of influence on other users.

However, it can be observed that the choice of recommendation algorithm and hyper-parameters (number of neighbors in kNN and the number of latent factors in NMF) has a significant impact to the degree to which some users can exert influence on other users. For example, In MovieLens increasing the number of latent factors used in NMF results in greater influence by a small number users  (Figure \ref{fig:Influencers_ML_NMF}). In Xing using a small number of factors can nearly eliminate influential users, but increasing the number of factors eventually has a diminishing effect on influence values. On the other hand, in Figure \ref{fig:Influencers_BX_NMF} we see the reverse of what we have observed in MovieLens with influence values consistently decreasing with larger numbers of latent factors. 

These results suggest that when performing parameter selection for optimizing the accuracy recommendation algorithms, consideration should also be given to the impact of these parameters on the distribution and the power of influential users. This is particularly important if mitigating the influence of a small number of users is critical in maintaining system level considerations such as balance and fairness. Our results here (as well as later experiments) show that managing the impact of influential users requires careful tuning of the underlying recommendation algorithms in a given dataset.

While the long-tail user influence plots reveal the distribution of influence values across users, they do not shed any light on the magnitude of the impact on other users. Furthermore, beyond the influence exerted by an individual user, it is also interesting to measure the influence of a small number of top influential users as a group. 

To this end, we designed a second set of experiments to address RQ1: we computed the percentage of users influenced by a relatively small number of top-$k$ influential users. In this experiment we intended to show the percentage of users influenced by top-$k$ most influential users, where $k$ is typically a small fraction (less than 5\%) of all the users. Furthermore, we measured the actual impact of an influential user $u$ on another user $v$ as the degree of overlap between the items in $u$'s profile and the recommendation set for $v$. We say that $v$ was ``impacted'' by $u$ if this overlap is greater than or equal to a pre-specified percentage threshold. 

\begin{figure*}[ht]
\centering
\begin{subfigure}{.33\textwidth}
  \centering
  \includegraphics[width=1\linewidth,trim=4cm 9cm 5cm 9cm,clip]{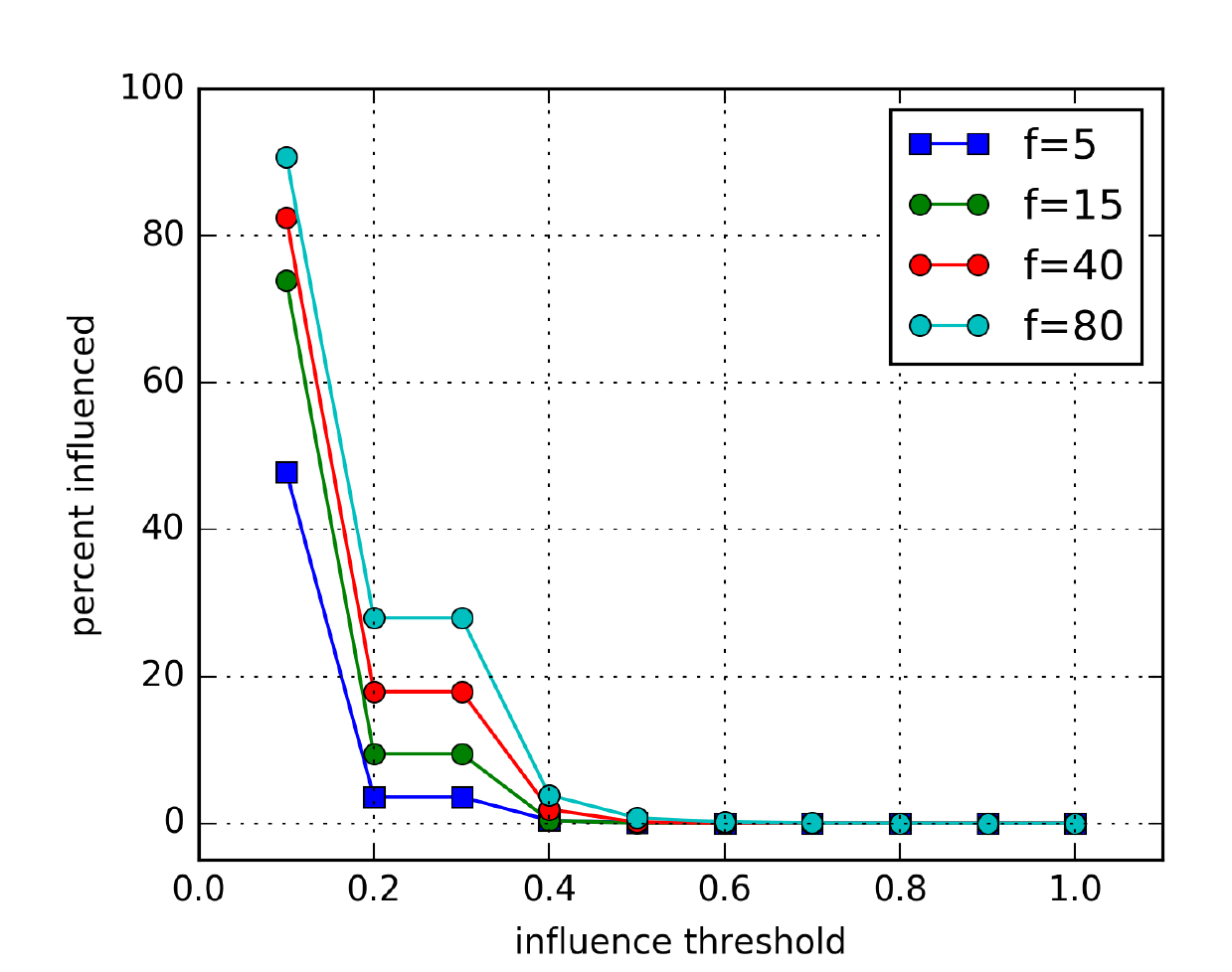}
  \caption{ML, influence of top-150 user}
  \label{fig:Infl_TH_ML_NMF}
\end{subfigure}%
\begin{subfigure}{.33\textwidth}
  \centering
  \includegraphics[width=1\linewidth,trim=4cm 9cm 5cm 9cm,clip]{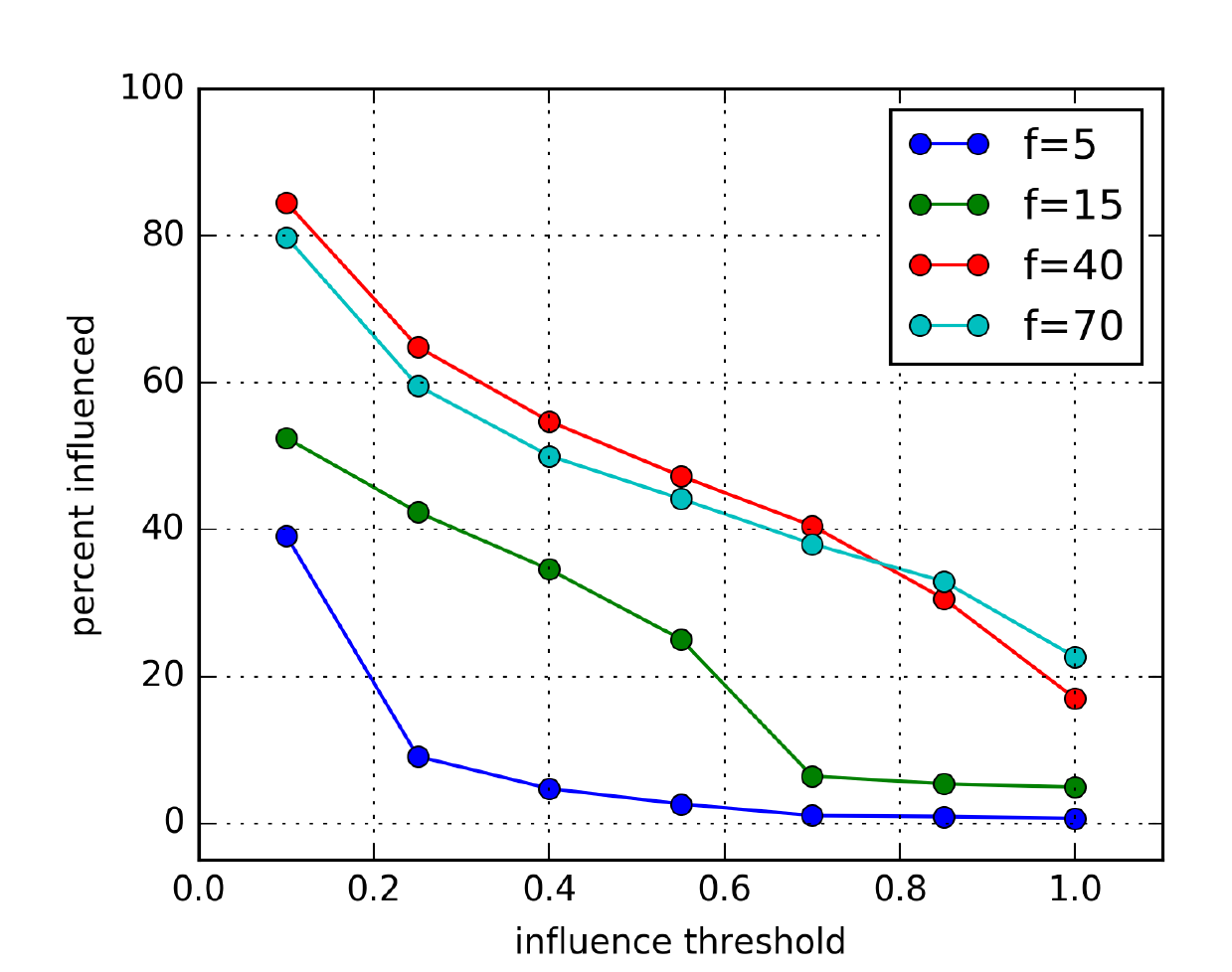}
  \caption{Xing, influence of top-50 user}
  \label{fig:Infl_TH_XG_NMF}
\end{subfigure}%
\begin{subfigure}{.33\textwidth}
  \centering
  \includegraphics[width=1\linewidth,trim=4cm 9cm 5cm 9cm,clip]{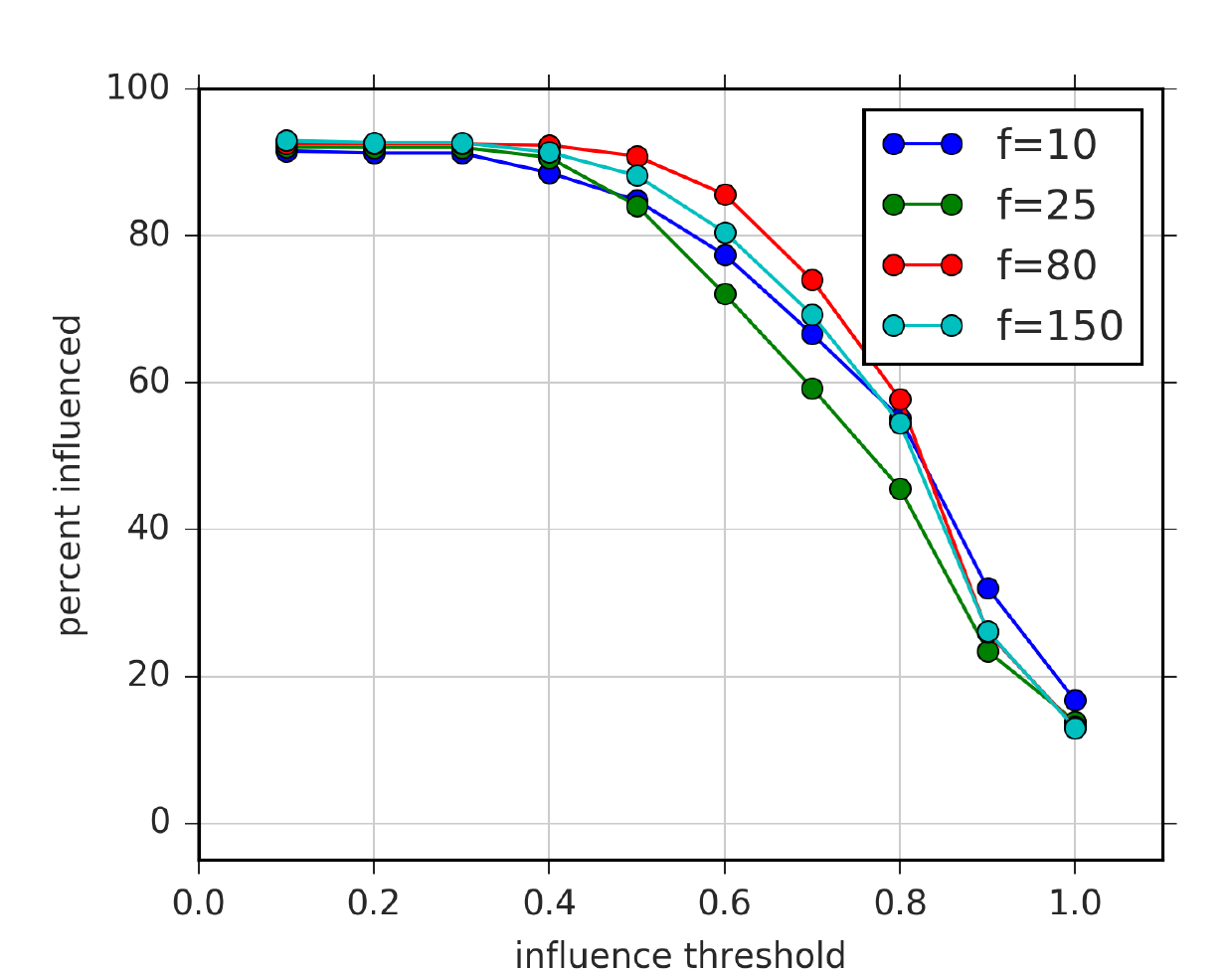}
  \caption{BX, influence of top-150 user}
  \label{fig:Infl_TH_BX_NMF}
\end{subfigure}%
\caption{Percentage of users influenced by top influencers in NMF's recommendations.}
\label{fig:influence_TH_NMF}
\end{figure*}

\begin{figure*}[ht]
\centering
\begin{subfigure}{.33\textwidth}
  \centering
  \includegraphics[width=1\linewidth,trim=4cm 9cm 5cm 9cm,clip]{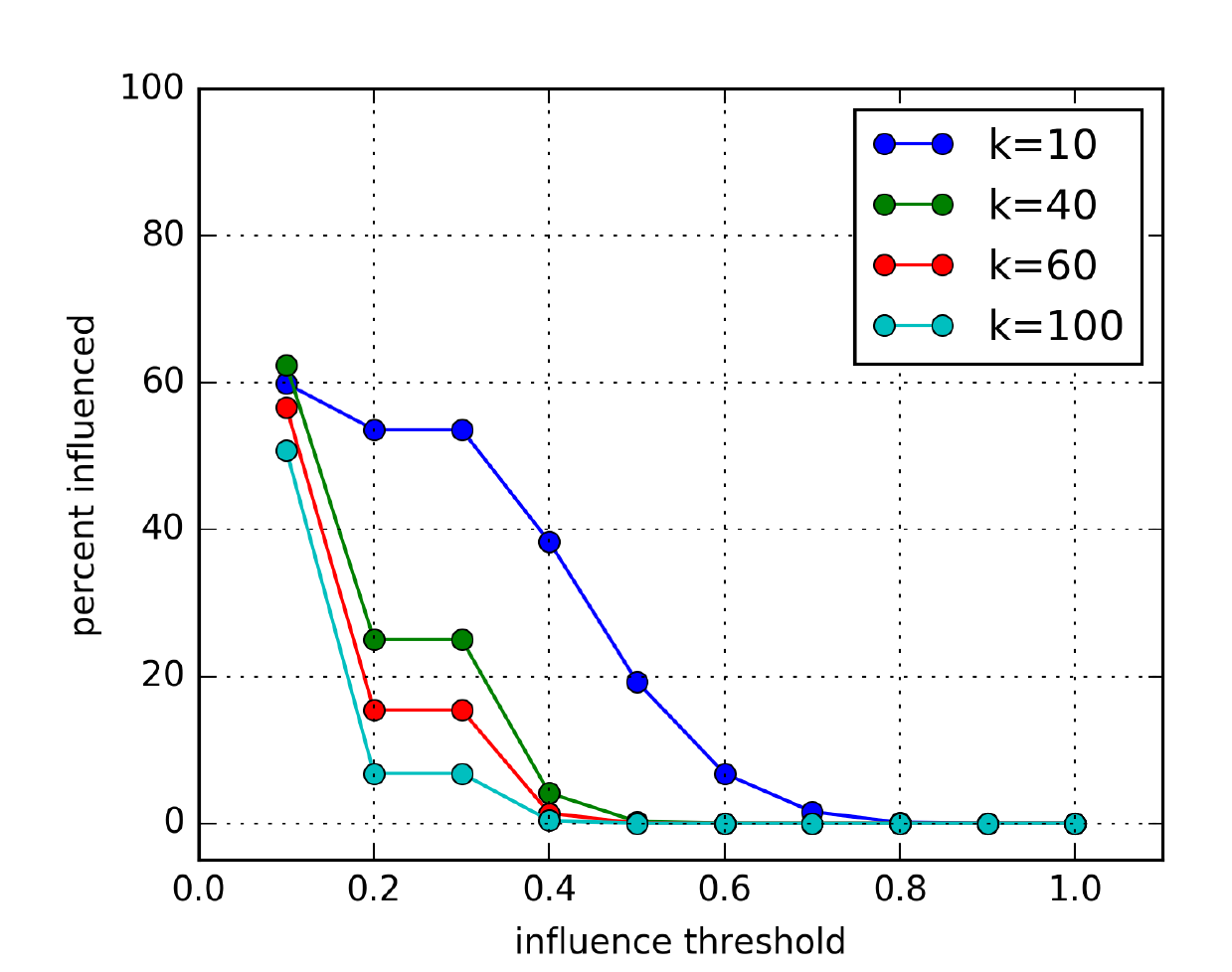}
  \caption{ML, influence of top-150 user}
  \label{fig:Infl_TH_ML_kNN}
\end{subfigure}%
\begin{subfigure}{.33\textwidth}
  \centering
  \includegraphics[width=1\linewidth,trim=4cm 9cm 5cm 9cm,clip]{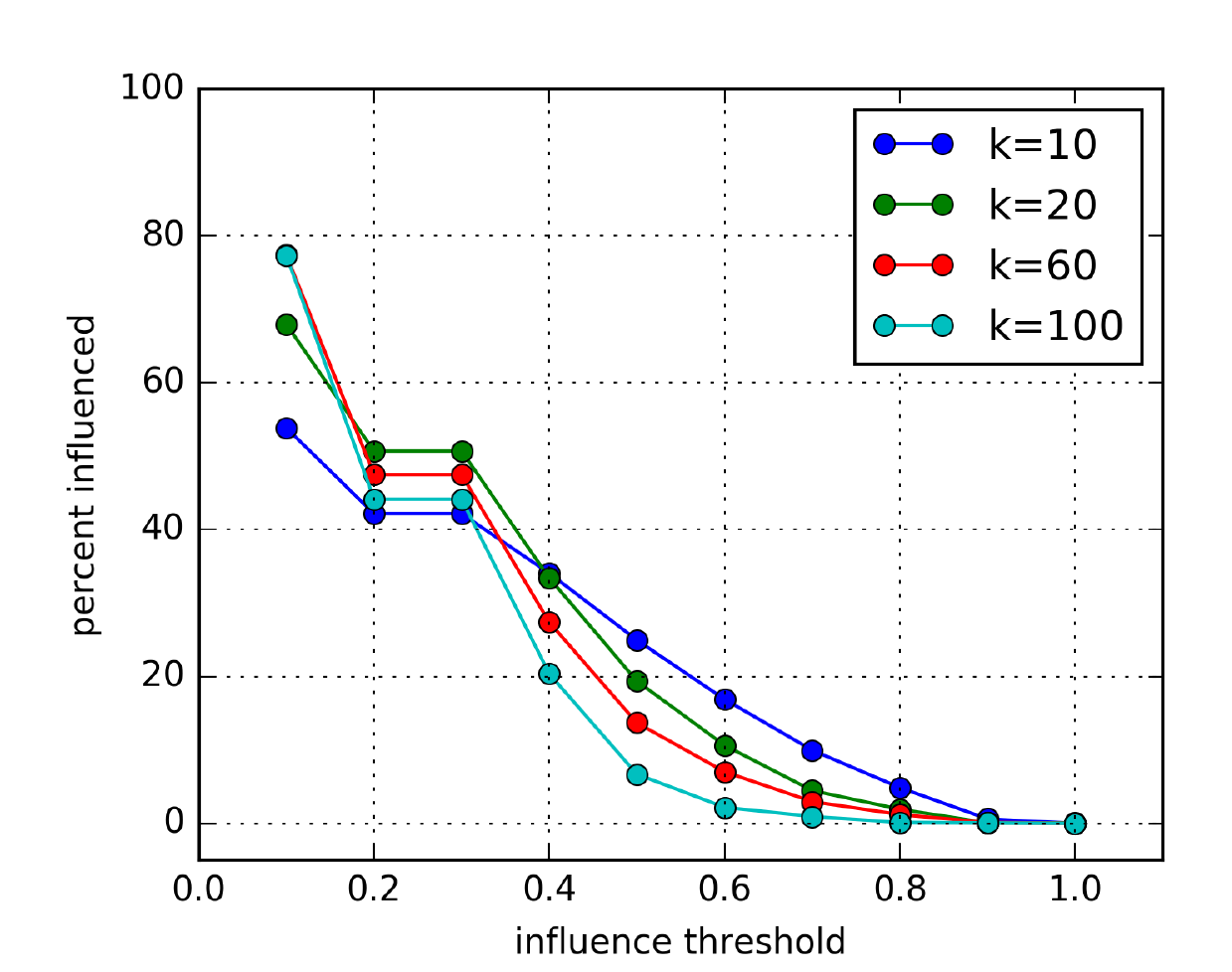}
  \caption{Xing, influence of top-50 user}
  \label{fig:Infl_TH_XG_kNN}
\end{subfigure}%
\begin{subfigure}{.33\textwidth}
  \centering
  \includegraphics[width=1\linewidth,trim=4cm 9cm 5cm 9cm,clip]{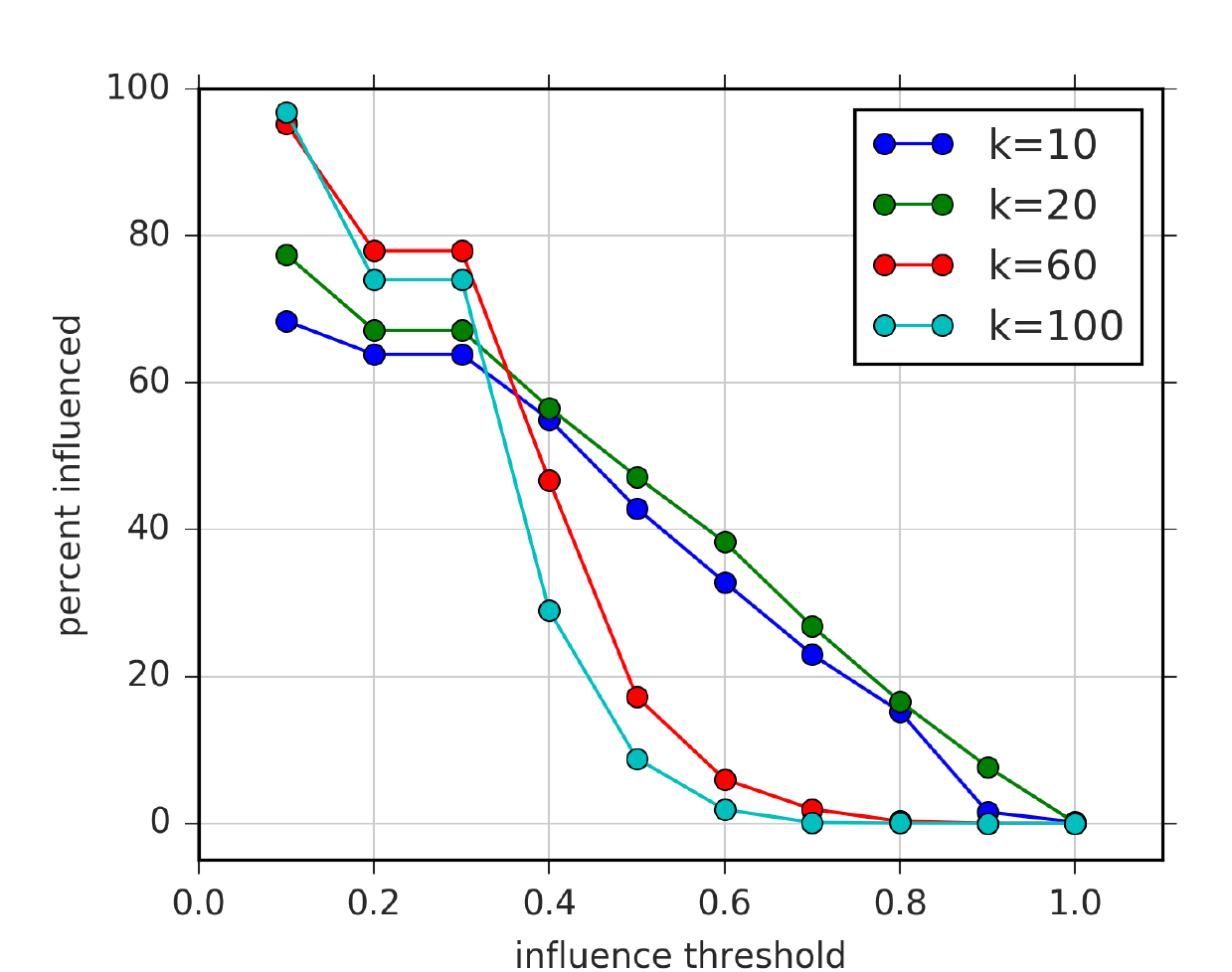}
  \caption{BX, influence of top-150 user}
  \label{fig:Infl_TH_BX_kNN}
\end{subfigure}%
\caption{Percentage of users influenced by top influencers in User-kNN's recommendations.}
\label{fig:influence_TH_kNN}
\end{figure*}

More formally, given a set of top influential users $\mathcal{T}$ and a threshold value $0 \leq \theta \leq 1$, the formula for the percentage influenced by top influential users is as follows:
\begin{equation}
    \sum_{u \in \mathcal{T}}{\hspace{0.3em} \sum_{v \in \{\mathcal{U} \setminus u\}}\bm{\mathds{1}}({Jaccard(\mathcal{R}_v^{(u)},  \hspace{0.3em} \overline{\mathcal{R}}_v^{(u)})} \geqslant \theta)}
\end{equation}
Where $Jaccard(R_1, R_2)$ is the normalized Jaccard distance of two recommendation sets $R_1, R_2$. Also, $\mathds{1}(x)$ is an indicator function which returns $1$ if the statement $x$ is true and returns zero otherwise. Note that the indicator function uniquely counts the users. In other words, when ${Jaccard(\mathcal{R}_v^{(u)}, \hspace{0.3em} \overline{\mathcal{R}}_v^{(u)})} \geqslant \theta$ is true for user $v$ and an influential user $u$ then it does not indicate user $v$ as $1$ (does not count it) again for any other influential user $u'$.

To focus on the impact of a small number of influential users as a group, we aggregated (taken the sum of) the influence values of top-$k$ influential users. In other words, we found the percentage of users who are influenced beyond a specified influence threshold by the top-$k$ influential users. 

These results are depicted in Figures \ref{fig:influence_TH_NMF} and \ref{fig:influence_TH_kNN}. To illustrate how these results can be interpreted consider the Xing plot for NMF with 70 latent factors ($f=70$). We see that 40\% of recommendations (influence threshold = 0.4) of about 55\% of users are changed due to the collective influence of the top-$50$ influential users. 

These result again support the conjecture that a small group of the most influential users can have a significant impact on the top recommendations provided to the rest of the users by the system. As in the case of previous experiment, we see that the choice of algorithm and their parameters have an impact on the size and scope of the influence exerted by influential users. For example, it seems that in general, matrix factorization results in a greater degree of impact by influential users on a larger percentage of users. This may be due to the fact that focusing on a small number of latent factors (instead of measuring similarities across large sets of items) amplifies the impact of influential users on recommendations. Further analysis is warranted to determine the exact nature of this effect.

Furthermore, the impact of influential users varies across data sets. It can be seen that the impact of influential users is much more muted in Movielens, while in Xing and (to an even larger degree) in Book Crossing (BX) influential users can have a much more dramatic impact on large proportions of users. We think this difference (e.g., in the BX data set) is probably due to the fact that some of these influential users are experts in reviewing books related to a set of prominent topics that may be captured by latent factors in NMF or commonly occur across user profiles in User-kNN. Further analysis of this effect is required in future work.

Once again these results attest to the impact a small group of influential users can have on the rest of the users. It also appears that this impact is generally larger and has greater scope in matrix factorization than in User-kNN. This choice of algorithm is critical in controlling the impact of influential users. 

Table \ref{tab:accuracy} shows the parameters of the most accurate models. Observing those accurate models in comparison to other models in figures \ref{fig:influencers_NMF}, \ref{fig:influencers_kNN}, \ref{fig:influence_TH_NMF}, and \ref{fig:influence_TH_kNN} suggests that there is no clear relation between ranking accuracy of recommendations and the dominance of influential users over other users. Therefore, traditional models in recommender systems besides accuracy, novelty and diversity measures \cite{ge2010beyondAcc, eskandanian2017clusteringDiv_evaluating}, can also benefit from Influence Discrimination model as an extra metric for the evaluation of recommendations, especially when balance and fairness are important considerations. 

\subsection{Low-Dimensional Visualization of Influential Users}
\label{sec:mds-plots}

In our third set of experiments, we tried to more directly address RQ2, i.e., to determine the similarities and relationships among influential users and between them and other users in the system. This is important because the main idea in collaborative filtering is to use like-minded users for generating recommendations. Thus the structural similarities among users in these systems become the core feature of their functionality. 

We used Multi-Dimensional Scaling (MDS) to visualize users (originally represented as vectors over items) in a 2-dimensional feature space. This is helpful in understanding the similarity structures among users and their relationships to influential users. MDS is useful for this type of analysis because it works by minimizing the total pair-wise distances of data points, which is exactly what we need for visualizing similarity structures among users. Highlighting the top influential users in a low-dimensional feature space would shed light on how they might influence similar users. In other words, the position of the influential users in 2D space relative to other users and each other would give us some insight into the relationship between profile similarity measures (such as Euclidean distance) and the degree of influence exerted.

\begin{figure}[h]
\centering
\begin{subfigure}{.46\textwidth}
  \centering
  \includegraphics[width=1\linewidth,trim=0cm 7cm 0cm 7cm,clip]{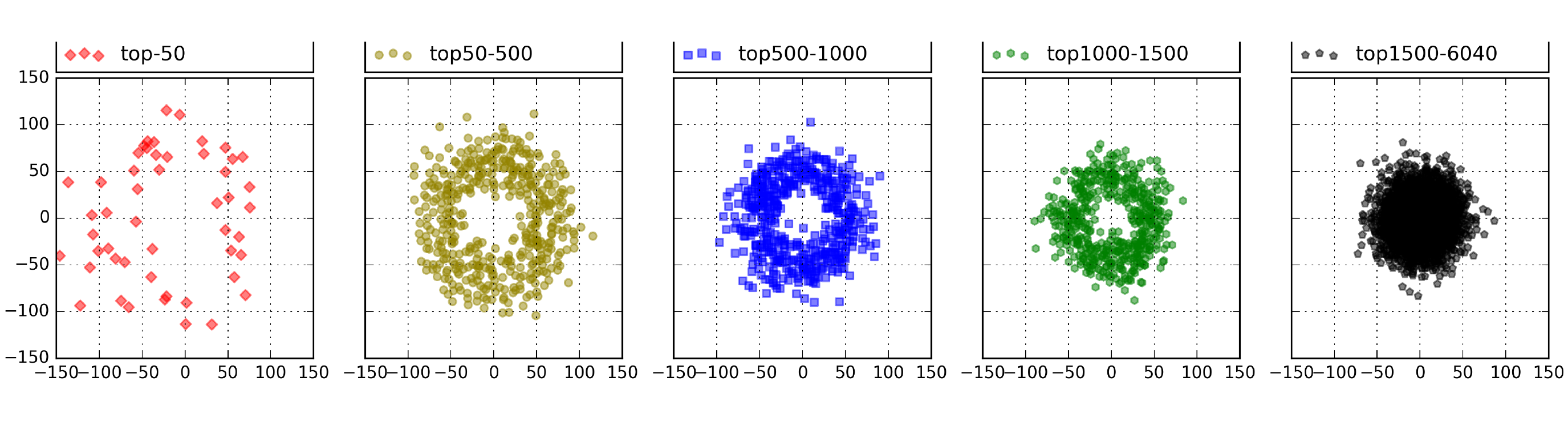}
  \caption{ML}
\end{subfigure}%

\begin{subfigure}{.46\textwidth}
  \centering
  \includegraphics[width=1\linewidth,trim=0cm 7cm 0cm 7cm,clip]{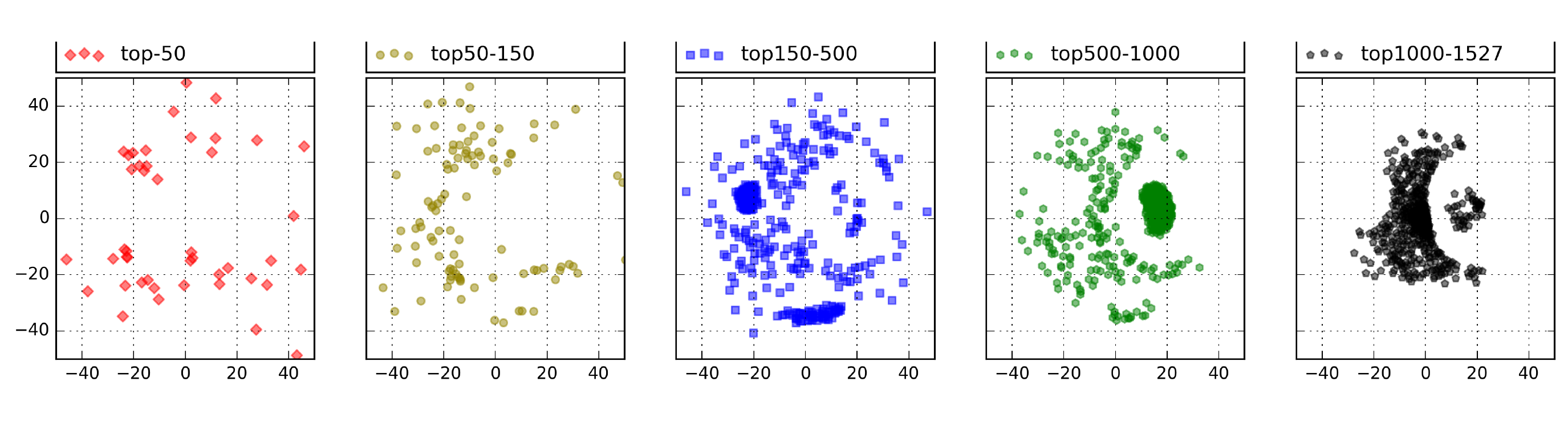}
  \caption{Xing}
\end{subfigure}%

\begin{subfigure}{.46\textwidth}
  \centering
  \includegraphics[width=1\linewidth,trim=0cm 7cm 0cm 7cm,clip]{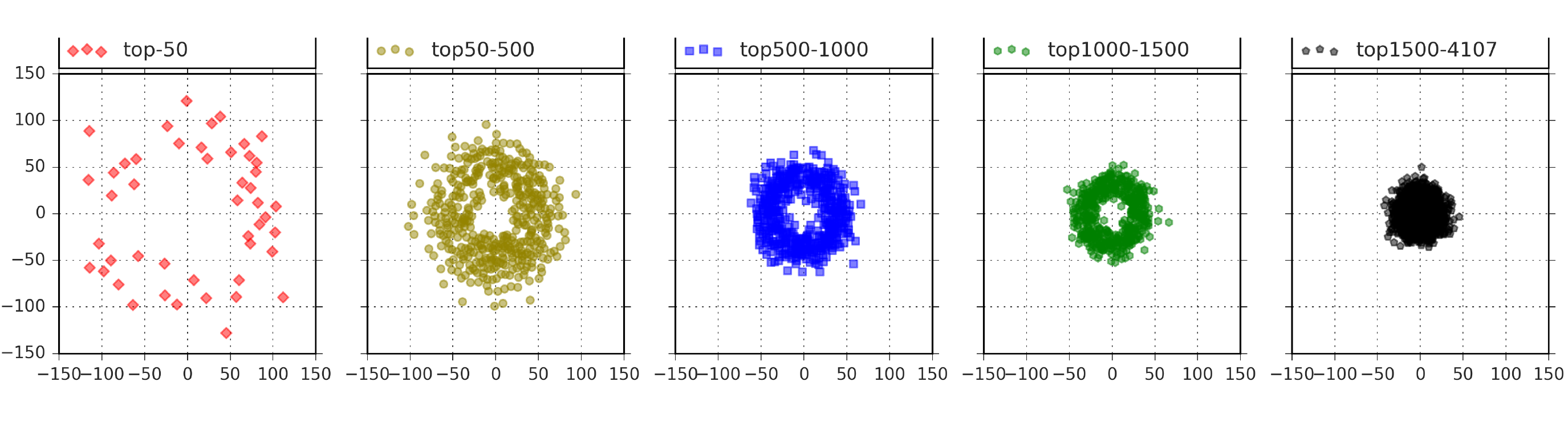}
  \caption{BX}
\end{subfigure}%
\caption{NMF results, MDS plots.}
\label{fig:nmf_mds_plots}
\end{figure}

\begin{figure}[h]
\centering
\begin{subfigure}{.46\textwidth}
  \centering
  \includegraphics[width=1\linewidth,trim=0cm 7cm 0cm 7cm,clip]{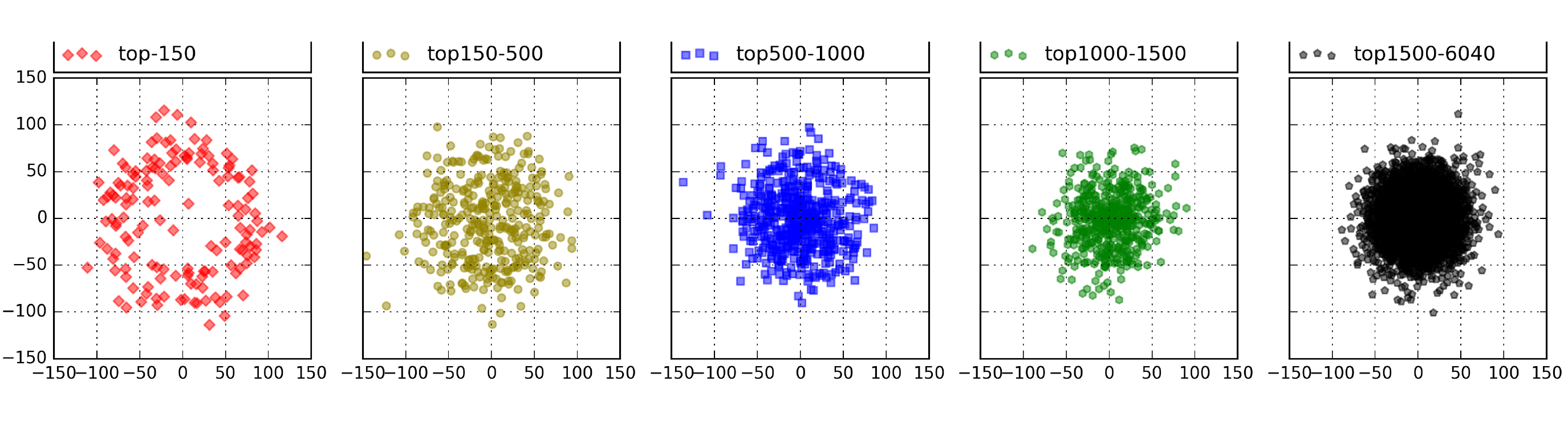}
  \caption{ML}
\end{subfigure}%

\begin{subfigure}{.46\textwidth}
  \centering
  \includegraphics[width=1\linewidth,trim=0cm 7cm 0cm 7cm,clip]{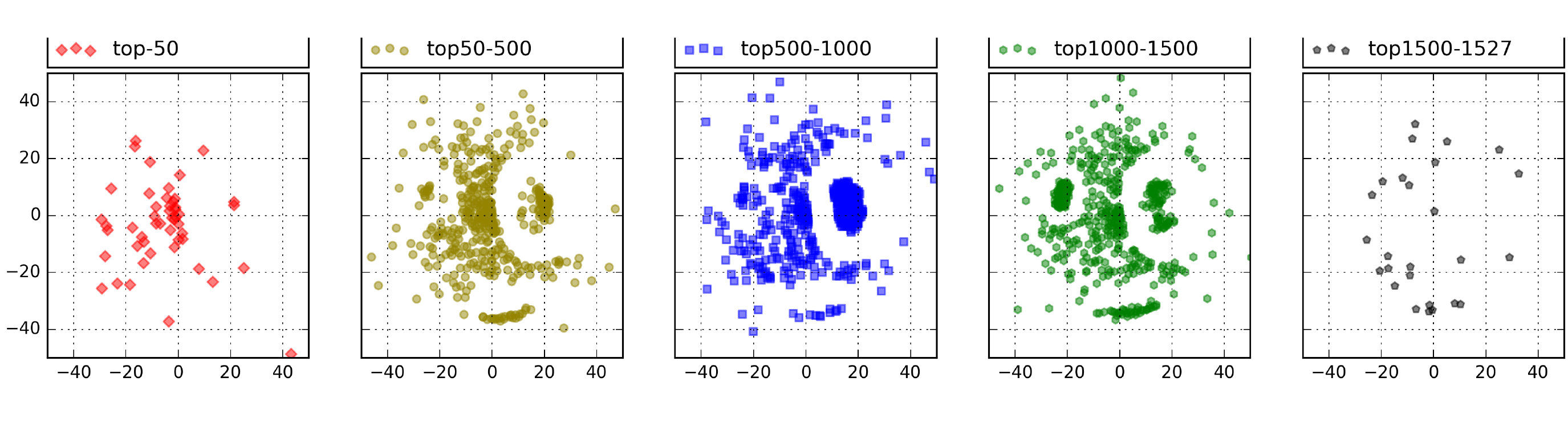}
  \caption{Xing}
\end{subfigure}%

\begin{subfigure}{.46\textwidth}
  \centering
  \includegraphics[width=1\linewidth,trim=0cm 7cm 0cm 7cm,clip]{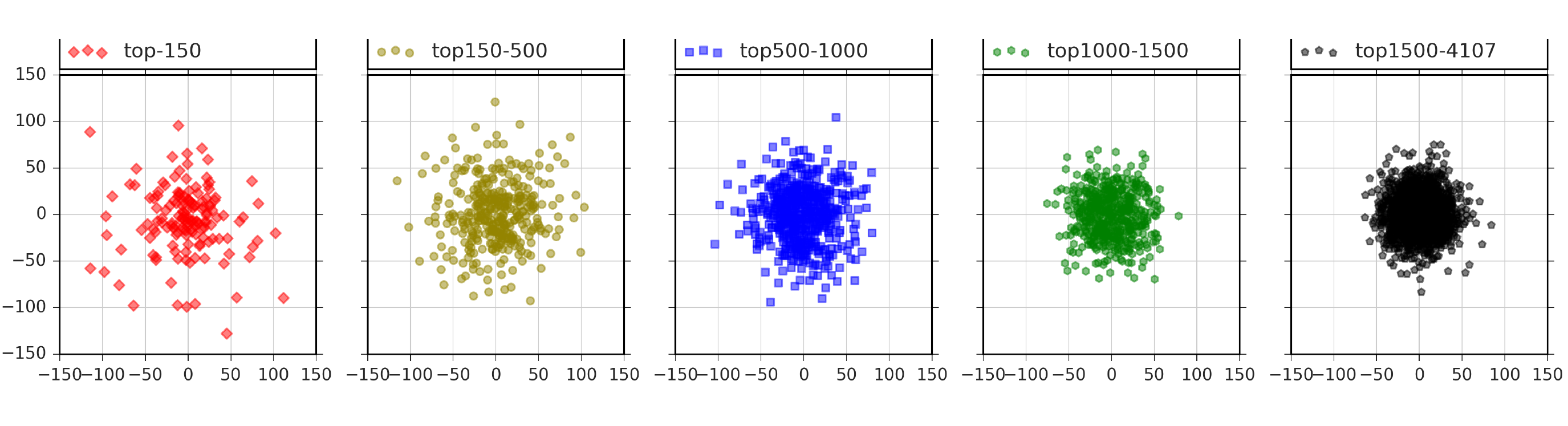}
  \caption{BX}
\end{subfigure}%
\caption{User-kNN results, MDS plots.}
\label{fig:knn_mds_plots}
\end{figure}

Figures \ref{fig:nmf_mds_plots} and \ref{fig:knn_mds_plots} show the density patterns among segments of influential users (with decreasing levels of influence by segments from left to right). So, the data points in the left most plots (red dots) in each dataset represent the few highly influential users (less than 5\% of all the users), and the right-most plot represents the largest segments of least influential users. The central point (coordinate (0, 0)) of these spaces are the most dense areas where most of the users are accumulated. Generally, the data points close to this center represent prototypical or average users in the system.

Interestingly, in the case of NMF (Figure \ref{fig:nmf_mds_plots}) almost all of the top influential users are located in the periphery, far away from the dense center. However, this pattern cannot be observed in User-kNN recommendations (Figure \ref{fig:knn_mds_plots} where the influential users are, in fact, more likely to be concentrated in the center. The most important finding in these figures is the difference in the density pattern of influential users in terms of their similarity structures between NMF and User-kNN. In NMF, influential users are relatively distant from majority of users and not very similar to each other. In User-KNN however, influential users tend to be more similar to each other and closer to the prototypical users in the system. We believe that this difference is due to the fact that in User-KNN, a user with the high profile similarity to many other users tends to contribute more to the recommendations made to these other users. In NMF, however, recommendations are generated based on user-item interactions across a small set of prominent latent factors. This reduces the reliance on user similarities across items and helps distinguish users with the most prominent presence of subsets of latent factors as influential users. In effect, the Influence Discrimination Model allows each algorithm to identify its own set of influential users with different characteristics than the other algorithms. We try to identify these different characteristics in the next set of experiments. But, further investigation using these and other algorithms is warranted and will be the subject of future work.

\subsection{Contributing Factors for Influence Estimation}

The main goal in our fourth set of experiments was to identify the features or characteristics that contribute the most to the degree of influence by a user. Identifying such features and determining their relative importance would help in uniquely characterizing the influential users. Furthermore, the Influence Discrimination Model is computationally expensive. In cases where millions of users are involved in generating recommendations it would not be feasible to use this model to measure influence of users. Therefore, it is useful to build an influence prediction model based on the available features about users, or at least find the most important features that can be used for this purpose. Earlier work in this area have studied heuristics based on various features to find the influential users. For example, inspired by metrics in social network analysis, authors in \cite{goyal2012recmax} have used the notion of user centrality (such as betweenness centrality) as a heuristic for estimating the influence of a user. Also, authors in \cite{rashid2005influence} have defined several characteristics of users according to their rated items such as the number of ratings, degree of agreements with others, rarity of rated items, and other features as indications of the degree of user influence. 


For our purposes, we use some of the same heuristics, based on user ratings, identified in this earlier work. But, we also identify some new factors that seem relevant based on our earlier analysis. Specifically, we consider the following factors as features for estimating the influence of a user and as the basis for characterizing influential users in a specific setting (e.g., using a specific recommendation algorithm).

\begin{itemize}
    \item $\bm{\beta_1}$: User $u$'s number of rated items $|I_u|$.
    
    \item $\bm{\beta_2}$: Centrality compared to other users, i.e., the average similarity of $u$ to all other users.
    
    \item $\bm{\beta_3$}: Number of times $u$ appears in the top-$k$ neighborhood of other users.
    $\beta_3(u) = \sum_{v \in \{\mathcal{U} \setminus u\}}{\sum_{u \in \mathcal{N}_v}{\mathds{1}}}$.
    
    \item $\bm{\beta_4}$: Given a distance threshold $\epsilon$ the number of users that find $u$ as their neighbor. In other words, this measure represents the number of users surrounding target user $u$ which is a measure of density of a user's neighborhood.
    
    $\beta_4(u) = \sum_{v \in \{\mathcal{U} \setminus u\}}{\sum_{u \in \mathcal{N}_{\epsilon}(v)}{\mathds{1}}}$,  
    where $N_{\epsilon}(v)$ $= \{v' \in \mathcal{U}$ $| \hspace{0.2em}$ $dist(v,v') < \epsilon \}$.
    
    \item $\bm{\beta_5}$: Average Jaccard similarity between rated items of u and recommendation lists of other users. 
    
    $\beta_5(u) = \frac{1}{|\mathcal{U}|-1} \sum_{v \in \mathcal{U}}{Jaccard(I_u, \mathcal{R}_v)}$.
    
    \item $\bm{\beta_6}$: Median popularity of items in $I_u$. Where the popularity of an item $i$ is defined by the number of times it has been rated:
    $Pop(i) = \sum_{v \in \mathcal{U}}{\sum_{i \in I_v}{\mathds{1}}}$.
    
    \item $\bm{\beta_7}$: Similarity of $u$ (as a vector of ratings) to the average user, where average user is defined as the centroid of all users (with mean rating for all items).
    
    \item $\bm{\beta_8}$: Intra-list distance of items in $I_u$. In other words, the average pair-wise distance of all the rated items by $u$ \cite{eskandanian2016Diversity}.
    
\end{itemize}

\begin{table*}[ht]
  \centering
  \begin{tabular}{c c | c c c c c c c c | c c}
    Rec& Dataset& $\beta_1$ & $\beta_2$ & $\beta_3$ & $\beta_4$ & $\beta_5$ & $\beta_6$ & $\beta_7$ & $\beta_8$ & $R^2$ & MSE \\ [0.5ex]
    \hline
    NMF& ML& 0.03 & 0.21 & 0.02 & \textbf{0.61} & 0.01 & 0.03 & 0.07 & 0.02 & 0.84 & 244.7 \\
    NMF& Xing& 0.01 & 0.25 & 0.02 & \textbf{0.33} & 0.19 & 0.13 & 0.05 & 0.00 & 0.80 & 210.1 \\
    NMF& BX& 0.01 & 0.31 & 0.02 & \textbf{0.52} & 0.07 & 0.04 & 0.01 & 0.01 & 0.95 & 436.5\\
    \hline
    kNN& ML& 0.01 & 0.00 & \textbf{0.95} & 0.00 & 0.01 & 0.00 & 0.05 & 0.00 & 0.96 & 2.11 \\
    kNN& Xing& 0.06 & 0.04 & \textbf{0.51} & 0.01 & 0.08 & 0.20 & 0.07 & 0.03 & 0.91 & 0.81 \\
    kNN& BX& 0.02 & 0.06 & \textbf{0.72} & 0.00 & 0.12 & 0.02 & 0.05 & 0.01 & 1.00 & 0.00 \\
    \hline 
  \end{tabular}
  \caption{Regression Tree results. $\beta_2$ (centrality), $\beta_3$ (number of times $u$ selected as the top neighbor of other users), and $\beta_4$ (density of users around $u$) are the most important factors.}
  \label{tab:regression}
\end{table*}
\begin{figure}[ht]
\centering
\begin{subfigure}[t]{.5\textwidth}
  \centering
  \includegraphics[width=\textwidth,trim=0cm 12cm 0cm 12cm,clip]{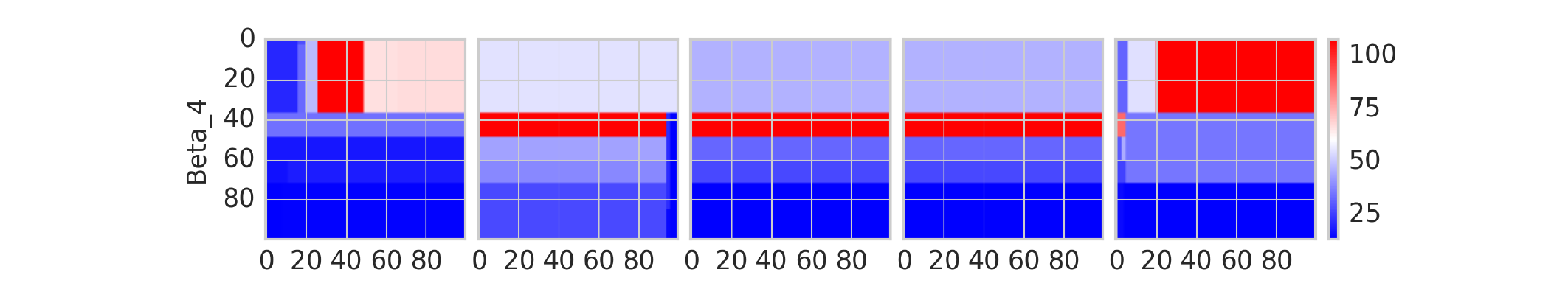}
  \caption{ML}
\end{subfigure}%

\begin{subfigure}[t]{.5\textwidth}
  \centering
  \raisebox{-\height}{\includegraphics[width=\textwidth,trim=0cm 12cm 0cm 12cm,clip]{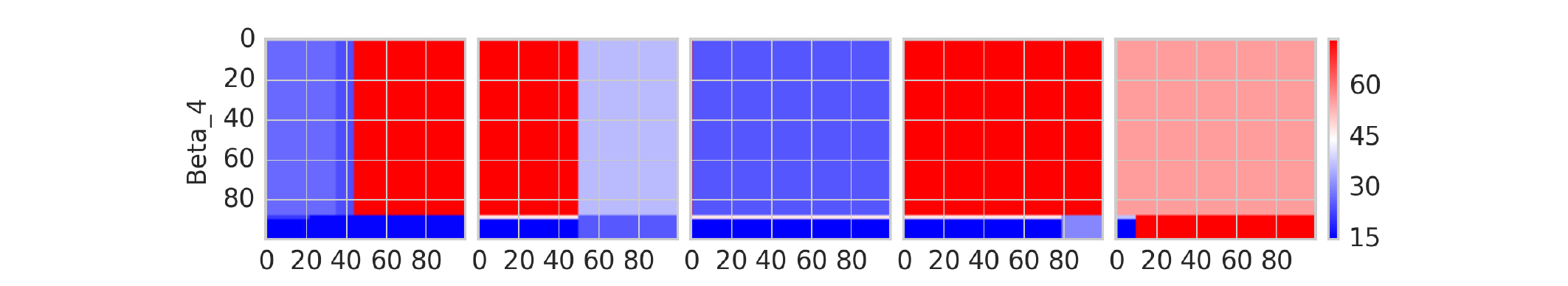}}
  \caption{Xing}
\end{subfigure}%

\begin{subfigure}[t]{.5\textwidth}
  \centering
  \raisebox{-\height}{\includegraphics[width=\textwidth,trim=0cm 12cm 0cm 12cm,clip]{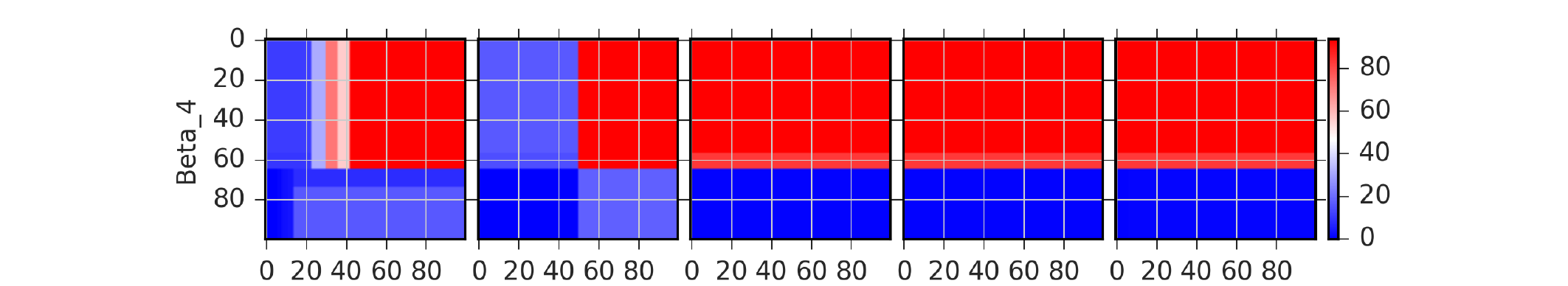}}
  \caption{BX}
\end{subfigure}%
\caption{NMF results, important features from left to right.}
\label{fig:feature_imp_nmf_plots}
\end{figure}

\begin{figure}[ht]
\centering
\begin{subfigure}{.5\textwidth}
  \centering
  \includegraphics[width=1\linewidth,trim=0cm 12cm 0cm 12cm,clip]{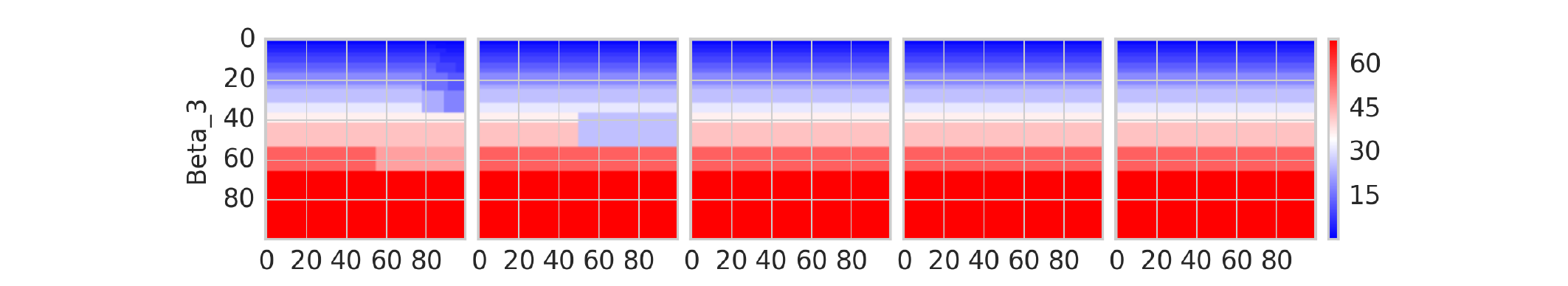}
  \caption{ML}
\end{subfigure}%

\begin{subfigure}{.5\textwidth}
  \centering
  \includegraphics[width=1\linewidth,trim=0cm 12cm 0cm 12cm,clip]{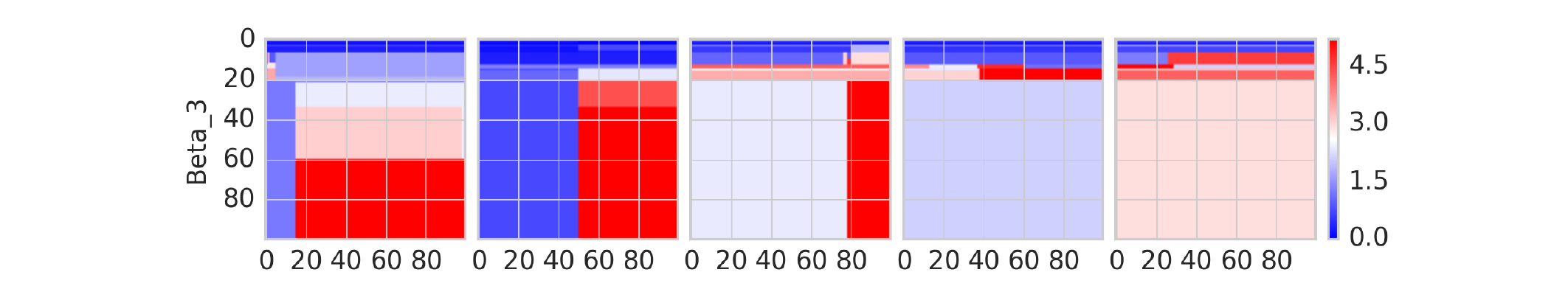}
  \caption{Xing}
\end{subfigure}%

\begin{subfigure}{.5\textwidth}
  \centering
  \includegraphics[width=1\linewidth,trim=0cm 12cm 0cm 12cm,clip]{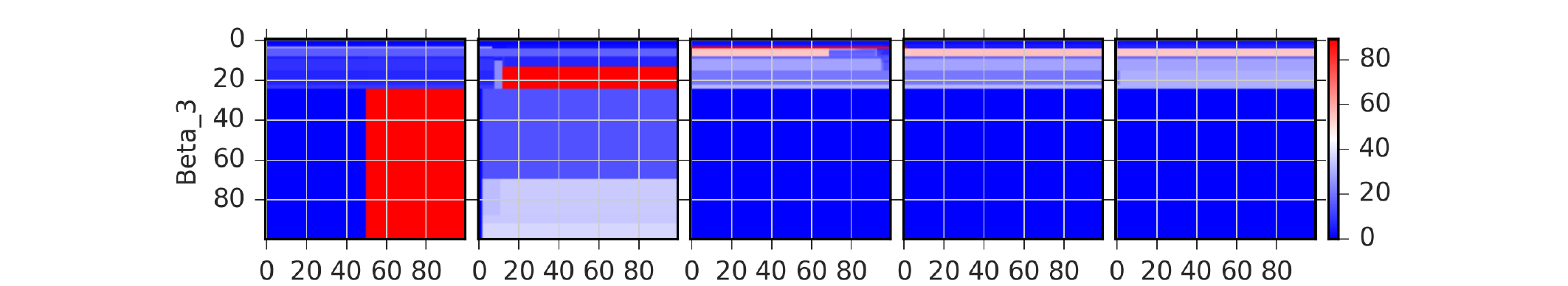}
  \caption{BX}
\end{subfigure}%
\caption{User-kNN results, important features from left to right.}
\label{fig:feature_imp_knn_plots}
\end{figure}

Using these features we aimed to build an influence prediction model. We used regression tree models for this purpose which are effective at handling nonlinear patterns in the data. These models generated the most accurate results compared to other regression models we tried. We kept the depth of these trees small (less than 10) so that they do not overfit data and can be generalized for unseen users. Table \ref{tab:regression} shows the result of these models. In regression trees the criterion which is used for splitting a node is mean squared error (MSE). The importance of a feature is computed as the normalized total reduction of the criterion resulting from that feature. Using this setting the importance of features are listed in table \ref{tab:regression}. 

NMF's results in this table suggest that $\beta_4$ and, to a lesser degree,  $\beta_2$ are the most important factors for predicting the influence of a user. Note that $\beta_4$ represents the density of users around a target user $u$, and $\beta_2$ is about betweenness centrality of $u$ in terms of user-user similarities. On the other hand, in kNN, $\beta_3$ is the most contributing factor for influence prediction. Although, the importance of $\beta_3$ in KNN is not surprising, the low importance of $\beta_4$ in kNN compared to NMF is interesting. The difference observed between KNN and NMF in these results are consistent with the observations we made about similarity structures in the MDS plots of Section \ref{sec:mds-plots}.

Although, table \ref{tab:regression} shows the importance of each factor, it does not reflect whether these correlations are positive or negative. We address this issue in Figures \ref{fig:feature_imp_nmf_plots} and \ref{fig:feature_imp_knn_plots} by visualizing the influence of decision boundaries using heatmaps. The decision boundaries of the most important features are selected and are shown in these figures. For example, in the MovieLens heatmaps in Figure \ref{fig:feature_imp_nmf_plots}, the red areas show the beta values for the high influence users. In this Figure the first heatmap from the left shows that the mid values of $\beta_2$ on the x axes and the values of $\beta_4$ on y-axes predict the highest influence values. This suggests a user $u$ who is moderately central ($\beta2$) and located in less dense areas ($\beta_4$) is highly influential (assuming that all the other betas are fixed at small values). 

Overall, these results suggest that the factors that characterize user influence are different given different recommendation algorithms. However, it is possible to build relatively accurate influence prediction models using a small set of factors for each algorithm.

\section{Conclusions and Future Work}

In this paper we introduced an \textit{Influence Discrimination Model} for more accurately measuring the influence of users in any Collaborative Filtering recommendation regardless of specific characteristics of the data and the underlying algorithms.
    
Using our Influence measure, we analyzed the behavior and the impact of most influential users across three real world datasets and using two different recommendation methods. We showed that influential users have different characteristics and impact other users differently in matrix factorization than in neighborhood-based methods. In general, the impact and the scope of influence of these users in matrix factorization is greater than kNN-based methods. However, our results show that the choice of parameters has an impact on the degree of influence exerted. So, algorithms must be tuned not only based on accuracy, but also by considering the impact of influential users on recommendations.
    
We identified several factors that can be used to characterize influential users, and using regression tree we built models for influence prediction. Results of these models suggest that influential users in matrix factorization have different characteristics than those in kNN-based methods.

In future work we will further explore the differences in characteristics of influential users among different algorithms and in different data sets. We will also extend our work to other recommendation approaches. 

\newpage
\bibliographystyle{ACM-Reference-Format}
\bibliography{ref}

\end{document}